\documentclass{PoS}
\usepackage{amssymb,here}

\title{First results of EPOS-HQ model for open heavy flavor production in A-A collission at RHIC and LHC}

\ShortTitle{First results of EPOS-HQ model for open heavy flavor production in AA at RHIC and LHC}

\author{\speaker{Pol Bernard Gossiaux}
	\\
SUBATECH Laboratory, IN2P3/CNRS, IMT Atlantique, Université de Nantes\\
E-mail: \email{gossiaux@subatech.in2p3.fr}}

\author{Joerg Aichelin\\
SUBATECH Laboratory, IN2P3/CNRS, IMT Atlantique, Université de Nantes\\
E-mail: \email{aichelin@subatech.in2p3.fr}}

\author{Benjamin Guiot\\
	Santa Maria University\\
	E-mail: \email{benjamin.guiot@usm.cl}}

\author{Iurii Karpenko\\
SUBATECH Laboratory, IN2P3/CNRS, IMT Atlantique, Université de Nantes\\
	E-mail: \email{karpenko@subatech.in2p3.fr}}

\author{Vitalii Ozvenchuk\\
	Institute of Nuclear Physics Polish Academy of Sciences, PL-31342 Krakow\\
	E-mail: \email{Vitalii.Ozvenchuk@ifj.edu.pl}}

\author{Tanguy Pierog\\
		KIT, Karlsruhe, IKP\\
	E-mail: \email{tanguy.pierog@kit.edu}}

\author{Jan Steinheimer\\
	Frankfurt University, FIAS\\
	E-mail: \email{steinheimer@fias.uni-frankfurt.de}}

\author{Klaus Werner\\
	SUBATECH Laboratory, IN2P3/CNRS, IMT Atlantique, Université de Nantes\\
	E-mail: \email{werner@subatech.in2p3.fr}}

\abstract{The EPOS-HQ model is a framework designed for predicting heavy-flavor observables both in p-p, p-A and A-A collisions. It results from the integration of the MC$@_s$HQ and EPOS3 codes and should supersede predictions made with each of these two. In this contribution, we   briefly present the new ingredients of the EPOS-HQ model as well as some preliminary results. From the comparison with experimental data, we derive some direction of improvement for the near future.}

\FullConference{International Conference on Hard and Electromagnetic Probes of High-Energy Nuclear Collisions\\
		30 September - 5 October 2018\\
		Aix-Les-Bains, Savoie, France}

\begin{document}
\section{Introduction}
While one of the ultimate goals of understanding heavy flavor production in ultrarelativistic pA and AA collisions is to assess the strength of heavy quark coupling with the dense phase created in such systems it is by now clearly established that "extra ingredients" such as the initial spectrum of heavy quarks or their hadronization mechanism play an important quantitative role in the predictions made for various observables such as the nuclear modification factor $R_{AA}$ or the elliptic flow $v_{2}$. In the past, we have suggested an energy loss model (MC$@_s$HQ) for HQ based on pQCD fledged with a running $\alpha_s$~\cite{Gossiaux:2008}, as the sum of both elastic and radiative processes. Predictions were made for RHIC~\cite{Gossiaux:2008} and LHC energies~(see \cite{Andronic:2016} and reference therein), with however different bulks and different initial distributions of HQ positions. In the later "bundle", MC$@_s$HQ was coupled to EPOS2, not appropriate for RHIC energies. In the meanwhile, the integration of MC$@_s$HQ and EPOS3 -- called EPOS-HQ -- was performed. We present significant evolutions from MC$@_s$HQ+EPOS2 $\rightarrow$ EPOS-HQ in section 2 and preliminary EPOS-HQ results in section 3.    
 
\section{The EPOS-HQ model}
\begin{figure}[H]
	\begin{center}
		\includegraphics[width=0.45\textwidth]{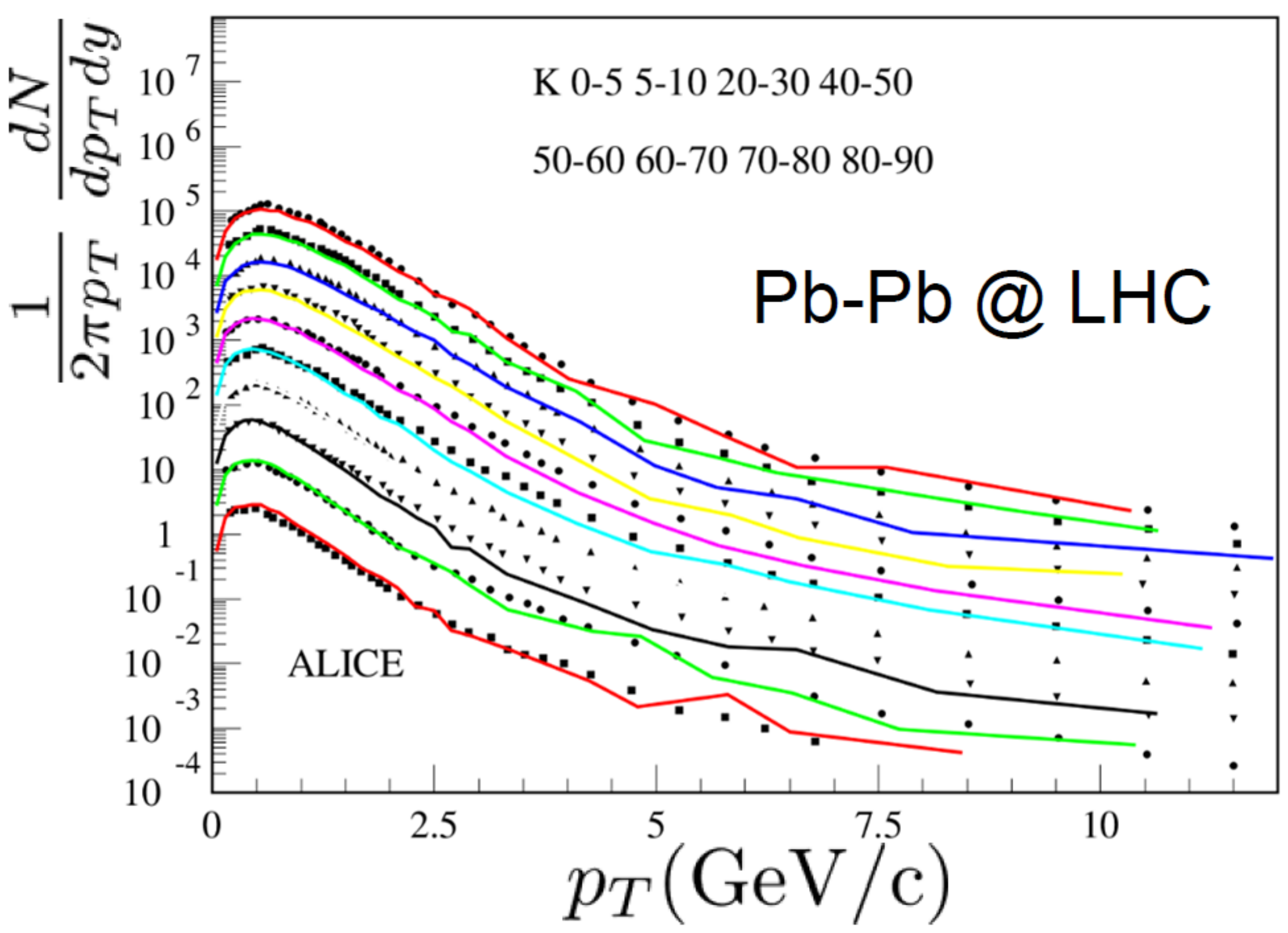}
		\includegraphics[width=0.45\textwidth]{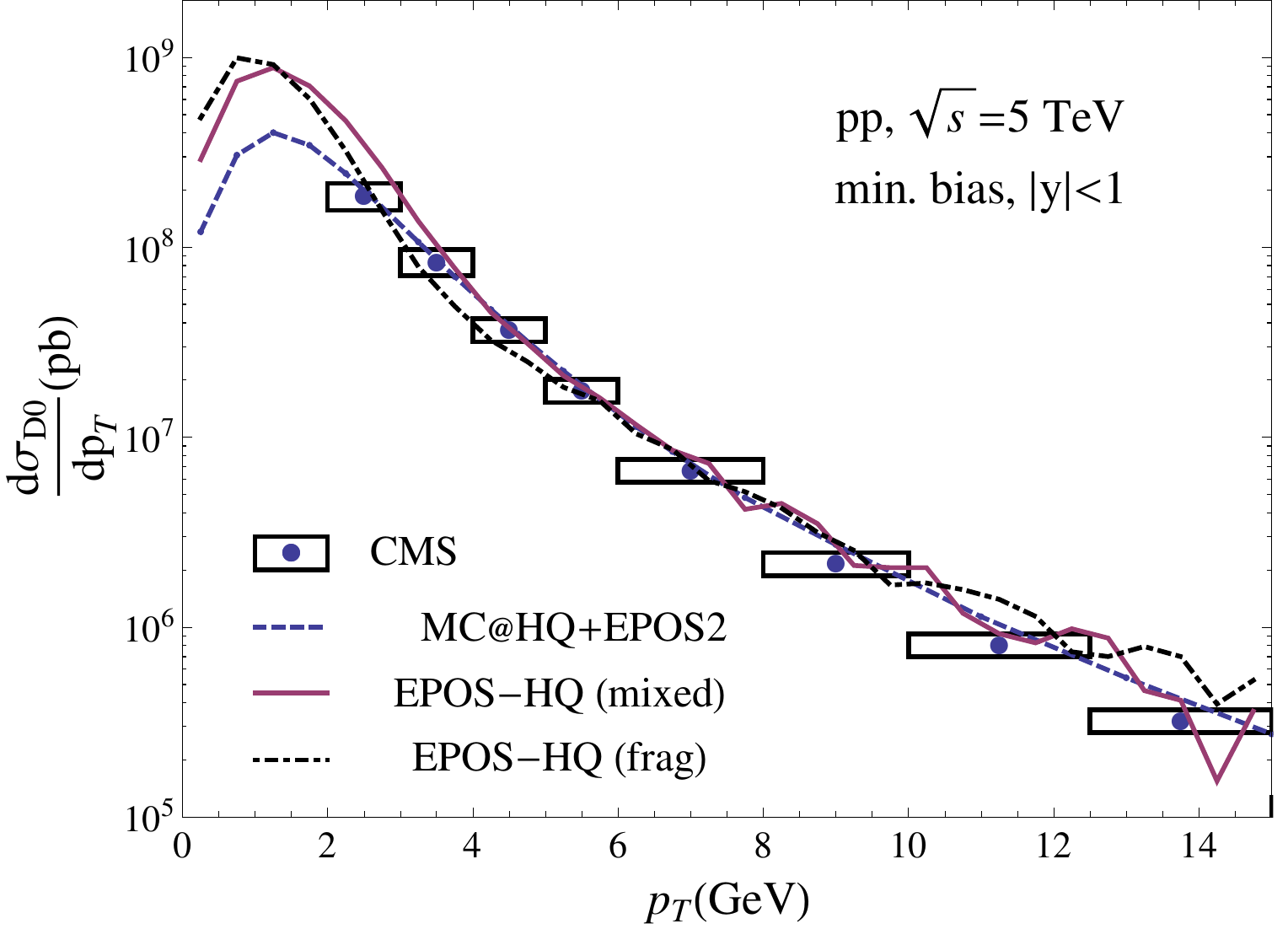}
	\end{center}
	\caption{Left: K production at LHC energy; comparison between EPOS3 and data. Right: $D_0$-meson $p_T$ spectra in pp from both MC$@_s$HQ and EPOS-HQ, compared to CMS data.}
	\label{fig1}
\end{figure}

While the energy loss model and the mixed (fragmentation+coalescence) hadronization mechanism is left unchanged as compared to the previous MC@sHQ+EPOS2 "bundle", the following components have evolved substantially in EPOS-HQ: a) the 3D+1 fluid dynamics vHLLE is upgraded in order to include viscous corrections \cite{Karpen:2014}, b) heavy quark production is now fully integrated as a part of the EPOS initial state, according to the semi-hard pomeron evolution scheme \cite{Guiot:2014}, including cold nuclear matter effects modeled through a saturation picture, 
c) final heavy flavor hadrons are propagated through the URQMD afterburner, where they can suffer elastic collisions.
On fig. \ref{fig1} (left panel) we 
illustrate the agreement reached in the soft sector by EPOS3. The $p_T$-spectrum of $D_0$ produced in $pp$ ($\sqrt{s}=5\,{\rm TeV}$) is displayed in the right panel, for the MC@sHQ model and for EPOS-HQ, where one can produce D mesons in pp either by c-quark fragmentation or through mixed hadronization process. It is important to notice that both choices lead to substantial differences with respect to the previous spectrum for $p_T\lesssim 4\,{\rm GeV}$ stemming from the initial c-quark distribution.

\section{Preliminary results and discussion}
\begin{figure}[H]
	\begin{center}
	\includegraphics[width=1\textwidth]{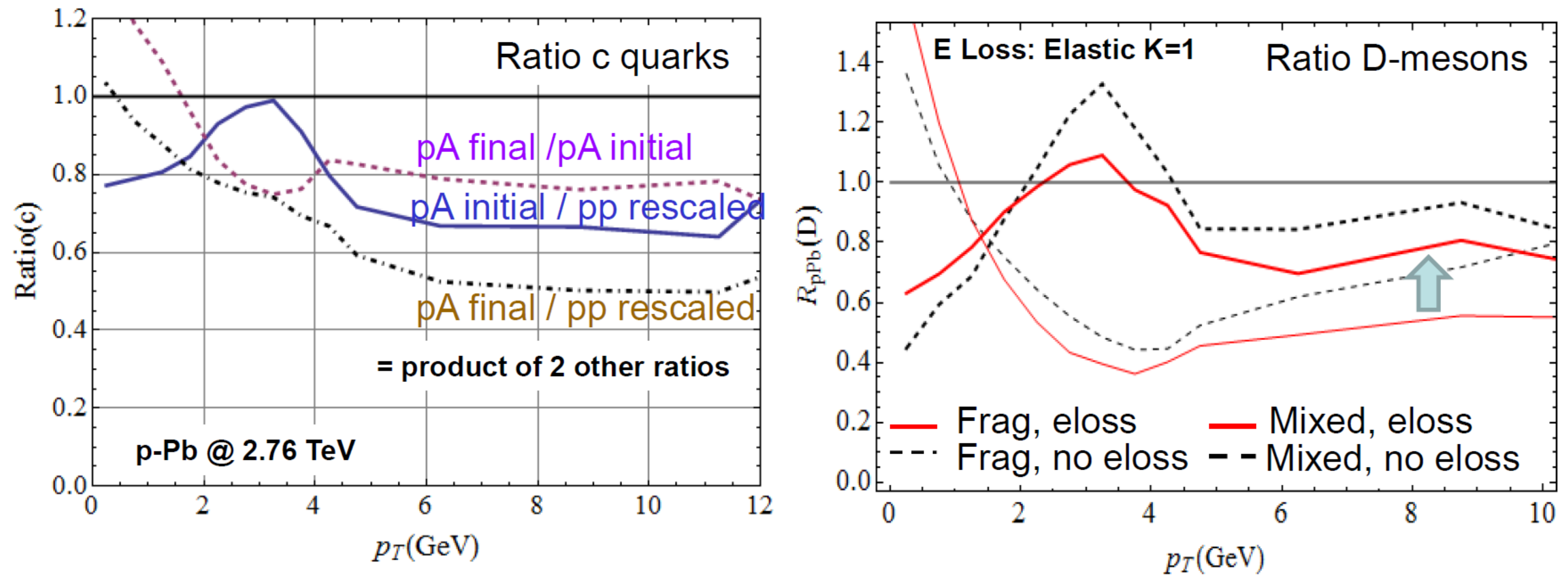}
	\end{center}
\vspace{-0.8cm}
	\caption{Left: ratio of various c-quark spectra at different stages of the evolution in p-Pb collisions. Right: corresponding $R_{\rm pPb}$ of D mesons for various options of EPOS-HQ.}
	\label{fig2}
\end{figure}

In this section, we show nuclear modification factors and $v_2$ of c quarks and D mesons for various systems and various options of the model. The p-Pb case at $\sqrt{s}=5\,{\rm TeV}$ (min. bias) is illustrated on fig. \ref{fig2}. In the present implementation of EPOS-HQ, cold nuclear matter (CNM) effects are still present at "large" $p_T$ (see full line on the left panel), resulting in $R_{pPb}(D)\approx 0.8$ for $p_T\gtrsim 5\,{\rm GeV}$. Quite generally, D-meson production can be understood as the convolution of several effects, including c-quark energy loss which is later on compensated when coalescence is considered as a possible hadronization mechanism (see right panel). Such observation is in clear tension with approaches including HF hadrons in the nPDF extraction procedure. 
\begin{figure}[H]
		\begin{center}
	\includegraphics[width=1.\textwidth]{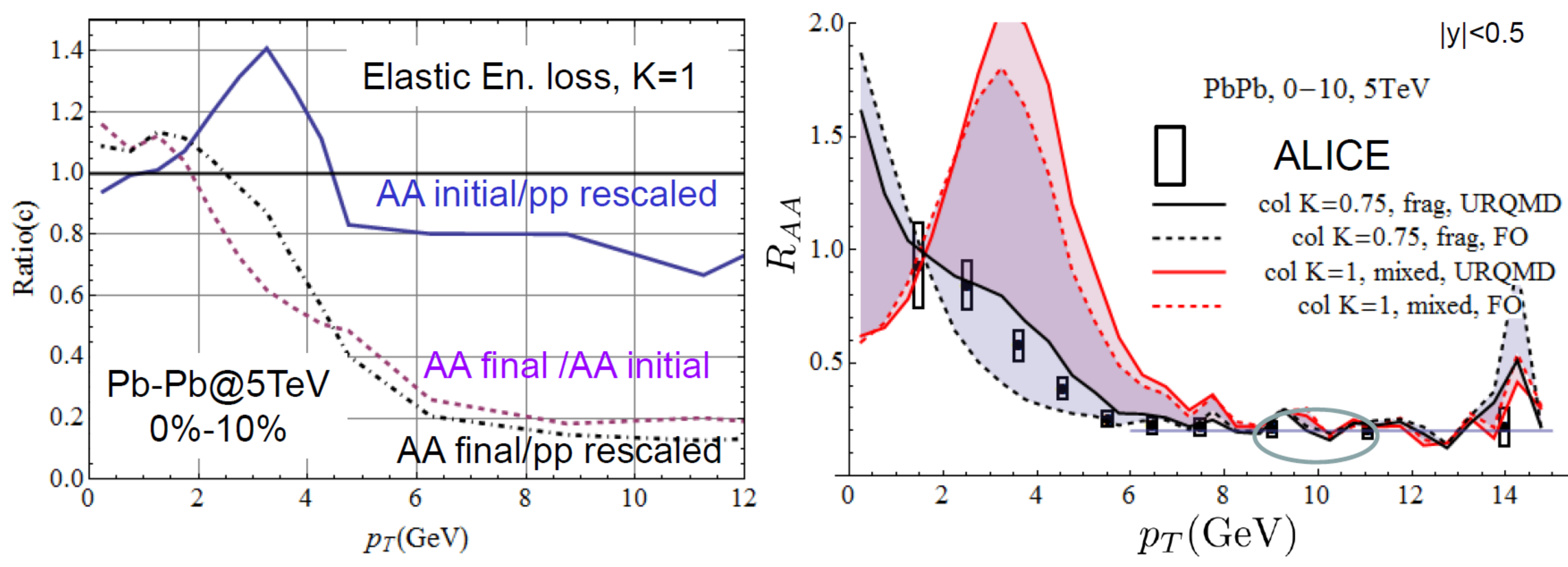}
		\end{center}
	\vspace{-0.8cm}
	\caption{Same as Fig. \ref{fig2} for central Pb-Pb collisions.}
	\label{fig3}
\end{figure}
On fig. \ref{fig3} (resp. fig. \ref{fig4}, left panel), we show results of our simulations for the Pb-Pb $R_{AA}(D)$ at $\sqrt{s}=5\,{\rm TeV}$, in the 0\%-10\% centrality class, with pure elastic (resp. elastic + radiative) energy loss calibrated on the experimental data at $p_T=10\,{\rm GeV}$ for both cases of pure fragmentation and mixed hadronization. On the right panel of fig. \ref{fig4}, the $v_2(D)$ for semi-central collisions is displayed. The most striking discrepancy with the experimental data is the pronounced bump observed around $p_T\approx 3\,{\rm GeV}$ if one resorts to mixed hadronization. While the precise hadronization procedure has a considerable impact on the D-meson spectrum in this $p_T$ range, producing those mesons through pure fragmentation is definitively disfavored by the $v_2$ analysis and contradictory to the paradigm which has recently emerged in the HF community. It should be noticed that such a bump was not predicted in our previous MC@sHQ+EPOS2 framework. A finer analysis reveals that the main ground for this evolution is the inital c-quark spectrum generated by EPOS3, much softer than our previous FONLL fit for $p_T\lesssim  5\,{\rm GeV}$. As a consequence, the evolution in the QGP do not lead to usual quenching in this region while the coalescence still generates an absolute $p_T$ increase, hence a harder spectrum for D mesons. 

\begin{figure}[H]
	\includegraphics[width=1.\textwidth]{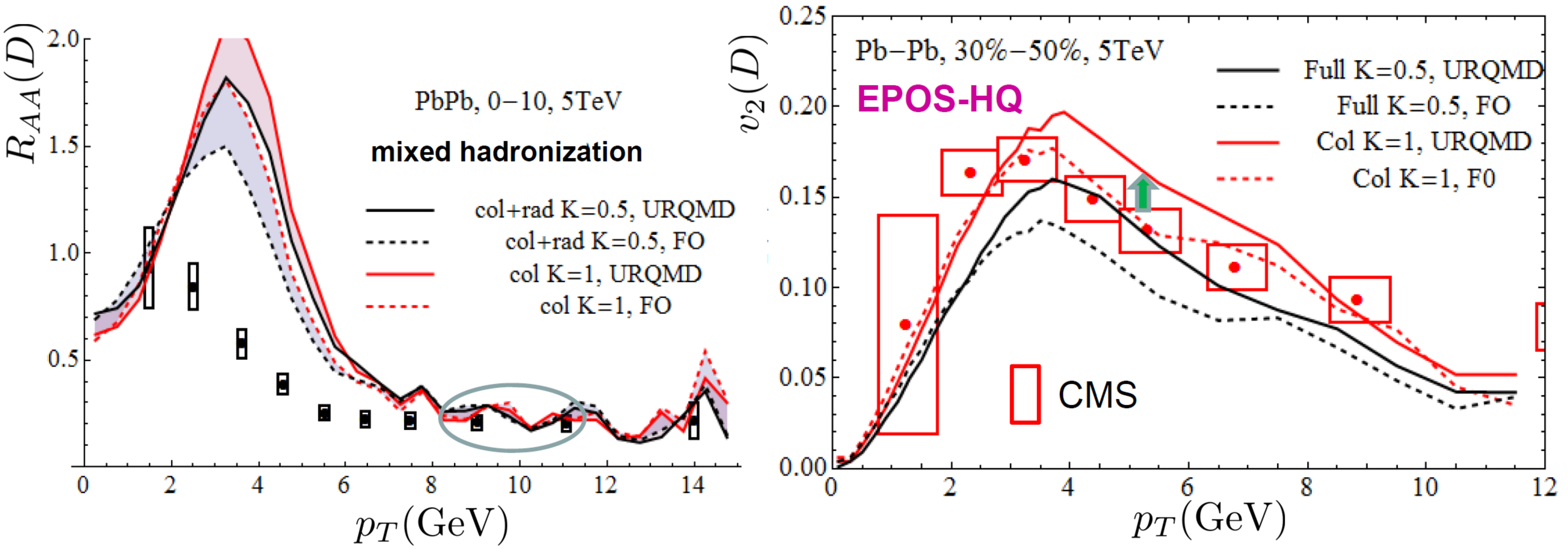}
		\vspace{-0.8cm}
	\caption{$R_{AA}$ (left) and $v_2$ (right) of D mesons produced in central Pb-Pb collisions for both types of energy loss, both of them being tuned on the ALICE $R_{AA}$ data at $p_T\approx 10\,{\rm GeV}$.}
	\label{fig4}
\end{figure}

\begin{figure}[H]
	\includegraphics[width=1.\textwidth]{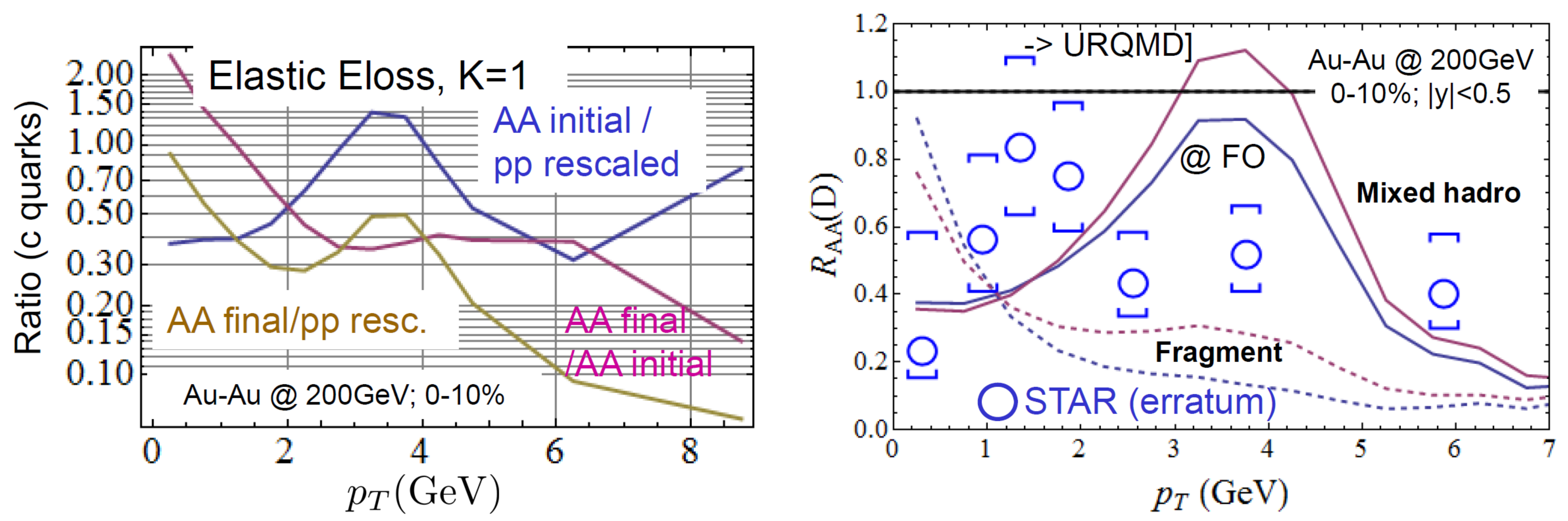}
		\vspace{-0.8cm}
	\caption{Same as fig. 3 for Au-Au collisions at $\sqrt{s}=200\,{\rm GeV}$, compared to the revised STAR data.}
	\label{fig5}
\end{figure}
While this feature obviously needs to be corrected in a near future, the good agreement with the experimental $v_2(D)$ -- less sensitive to the initial c-quark spectrum -- should be considered as a positive sign for the relevance of the global framework. The right panel of Fig. \ref{fig4} also illustrates the clear importance of the URQMD afterburner, which can lead up to a 3\% increase of the $v_2(D)$.

Similar observables are shown in fig. \ref{fig5} and \ref{fig6} (left panel) for Au-Au collisions at RHIC top energy, using the same modeling of the energy loss (with the same cranking coefficient $K$ of the elementary cross section). Pretty similar conclusions can be derived as for the LHC reactions: while the $v_2(D)$ is found in good agreement for both the pure elastic and the elastic + radiative energy loss. 
Reproducing the $R_{AA}$ appears to be out of the reach of our EPOS-HQ model in the present stage: If one indeed obtains a flow bump when resorting to mixed hadronization, it is positioned at too large $p_T$ as compared to the corrected STAR data; besides the initial CNM effects appear to be too pronounced in the intermediate $p_T$ range.

\section{Conclusions and perspectives}
\begin{figure}[H]
	\includegraphics[width=1.\textwidth]{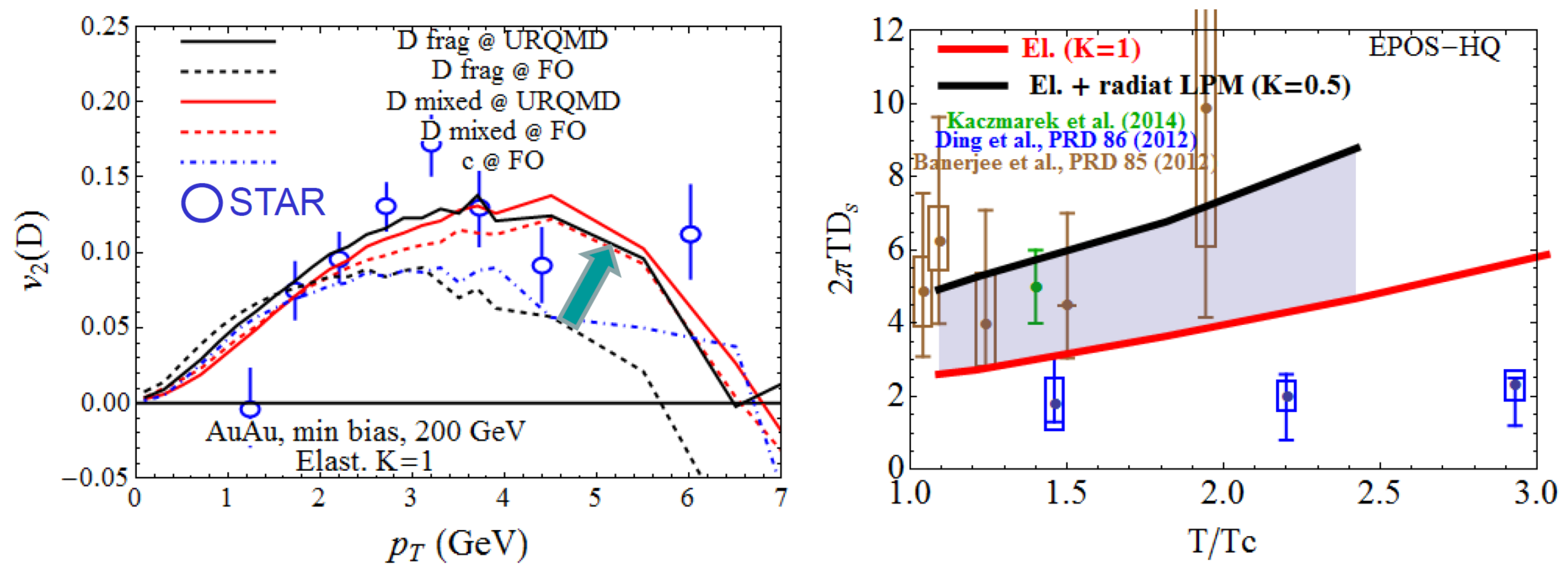}
		\vspace{-0.8cm}
	\caption{Left: $v_2(D)$ for both types of hadronization, with or without URQMD afterburner; the arrow highlights the contribution of the URQMD afterburner when hadronization proceeds through fragmentation, leading to unsaturated $v_2$. Right: Spatial diffusion coefficient resulting from the new EPOS-HQ calibration.}
	\label{fig6}
\end{figure}

We have presented preliminary results for open heavy flavor production in some p-A and A-A systems at RHIC and LHC. The evolution of some of some main ingredients of the framework leads to a new calibration of the energy loss model. The corresponding $D_s$ coefficient, presented in fig.~\ref{fig6} (right panel), are compatible with state of the art lQCD calculations and found to be 50\% larger as compared to values extracted from MC$@_s$HQ+EPOS2~\cite{Andronic:2016}. Although a good agreement was achieved for the $v_2(D)$ observable, being able to cope with the $R_{AA}$ requires some extra work, in particular as regards the initial c-quark spectrum in those systems.     

\begin{thebibliography}{99}
\bibitem{Gossiaux:2008}
P.B. Gossiaux and J. Aichelin, \emph{Toward an understanding of the RHIC single electron data}, \emph{Phys. Rev. C} {\bf 78} (2008), 014904. 

\bibitem{Andronic:2016}
A. Andronic et al.,\emph{Heavy-flavour and quarkonium production in the LHC era: from
	proton–proton to heavy-ion collisions}, \emph{Eur. Phys. J. C} {\bf 78} (2016), 107. 
	
\bibitem{Karpen:2014}
Iu. Karpenko, P. Huovinen, M. Bleicher, \emph{A 3+1 dimensional viscous hydrodynamic code for relativistic heavy ion collisions}, \emph{Comput. Phys. Commun.} {\bf 185} (2014), 3016.

\bibitem{Guiot:2014}
B. Guiot, Ph.D. Thesis, University of Nantes, (2014), 2014EMNA0192, tel-01127223.

\bibitem{Werner:2014}
K. Werner et al., \emph{Analysing radial flow features in p-Pb and p-p collisions at several TeV by studying identified particle production in EPOS3}, \emph{Phys.Rev. C} {\bf 89} (2014), 064903.

\bibitem{Lang:2012cx}
T.~Lang, H.~van Hees, J.~Steinheimer, G.~Inghirami and M.~Bleicher,
\emph{Heavy quark transport in heavy ion collisions at energies available
at the BNL Relativistic Heavy Ion Collider and at the CERN Large Hadron
Collider within the UrQMD hybrid model}, \emph{Phys.\ Rev.\ C} {\bf 93} (2016), 014901.

\end{thebibliography}
\end{document}